\documentstyle[12pt]{article}
\begin{document}
\def\thefootnote{\fnsymbol{footnote}}
\vspace*{2cm}
\begin{center}
{\LARGE\bf Leptogenesis and dark matter unified 
in a non-SUSY model for neutrino masses}\\
\vspace{1 cm}
{\Large Daijiro Suematsu}
\footnote{e-mail:~suematsu@hep.s.kanazawa-u.ac.jp}
\vspace {1cm}\\
{\it Institute for Theoretical Physics, Kanazawa University,\\
        Kanazawa 920-1192, Japan}\\
\end{center}
\vspace{1cm}
{\Large\bf Abstract}\\
We propose a unified explanation for the origin of dark matter and 
baryon number asymmetry on the basis of a non-supersymmetric model 
for neutrino masses. Neutrino masses are generated in two distinct 
ways, that is, a tree-level seesaw mechanism with a single right-handed 
neutrino and one-loop radiative effects by a new additional doublet scalar.
A spontaneously broken U(1)$^\prime$ brings a $Z_2$ symmetry which
restricts couplings of this new scalar and controls the neutrino
masses. It also guarantees the stability of a CDM candidate. 
We examine two possible candidates for the CDM.
We also show that the decay of a heavy right-handed neutrino related 
to the seesaw mechanism can generate baryon number asymmetry through 
leptogenesis.

\newpage
\setcounter{footnote}{0}
\def\thefootnote{\arabic{footnote}}
\section{Introduction}
Neutrino masses \cite{nmass}, cold dark matter (CDM) \cite{wmap}, 
and baryon number asymmetry in the universe \cite{basym} suggest 
that the standard model (SM) should be extended.
Both neutrino masses and baryon number asymmetry are well known to be 
explained in a unified way through the leptogenesis scenario 
in the framework of the seesaw mechanism \cite{leptg1}. 
Extensive studies have been done on this subject during recent 
several years \cite{leptg2}.
On the other hand, supersymmetry is known to play a crucial role for the
explanation of CDM abundance in the universe \cite{cdm1}, although it
has been introduced originally to solve the hierarchy problem.
Supersymmetric models have good candidates for CDM such as
the lightest superparticle (LSP) as long as $R$-parity is conserved.
The neutralino LSP has been extensively studied as a CDM candidate 
in the supersymmetric SM (MSSM) and its singlet extensions \cite{cdm2,cdm2e}.
If we try to explain simultaneously both the leptogenesis and 
the CDM abundance in supersymmetric models, we have a difficulty.
The out-of-equilibrium decay of thermal heavy neutrinos 
can generate sufficient baryon number asymmetry only if the reheating
temperature is high enough such as $T_R>10^8$~GeV. For such
reheating temperature, however, we confront the serious gravitino 
problem in supersymmetric models \cite{gravitino,gravitino2}.
Various trials to overcome this difficulty have been done by searching
scenarios to enhance the $CP$ asymmetry and lower the required reheating
temperature \cite{resonant,softlept1,sgravitino}.

In these studies, the CDM and the baryon number asymmetry 
are separately explained based on unrelated physics. Thus, 
we cannot expect to obtain any hints as to 
why the CDM abundance is of similar order 
as the baryon number asymmetry in the present universe through 
such studies.\footnote{There are 
several works to relate the CDM abundance to the baryon number
asymmetry. For such trials, see \cite{bdsol} for example.} 
Unfortunately, at present, we have no satisfactory supersymmetric 
models to explain these three
experimental evidences which impose us to extend the SM.
In this situation it may be worth to take a different empirical view
point at first and reconsider possible models which can explain these 
evidences simultaneously
on the basis of closely related physics \cite{extsm}. 
As the next step, the hierarchy problem may be considered in the 
framework where such models are embedded.

Recently, it has been suggested that neutrino masses 
and the CDM abundance may be related in some kind of non-supersymmetric 
models for neutrino masses.  In such models neutrino masses are generated 
through one-loop radiative effects which are induced by new scalar 
fields \cite{z2}.
A certain $Z_2$ symmetry prohibiting large neutrino masses can 
also guarantee the stability of a CDM candidate like $R$-parity in
supersymmetric models \cite{nonsusy1,nonsusy2,nonsusyz}. 
The baryon number asymmetry has also been discussed 
in this model \cite{nonsusyb}.
In the same type model there is also a suggestion that the hierarchy
problem can be improved by considering a heavy Higgs scalar \cite{bar}.  
Since these models have rather simple structure at weak scale regions, 
it might give us some useful hints for physics beyond the SM
if they can explain the above mentioned experimental evidences consistently.

In this paper, we consider the possibility that the baryon number
asymmetry is closely related to the origin of both neutrino 
masses and CDM abundance. We show that the ordinary leptogenesis 
based on heavy neutrino decay can be embedded consistently in the 
model for neutrino masses proposed in \cite{nonsusyz}. 
As we discuss below, this is closely related to an extension of
\cite{nonsusyz} such that (1) an additional $N$ with zero charge under 
U(1)$^\prime$ is introduced and (2) the dimension five term in the
scalar potential has a complex coupling $\lambda_6$.
The paper also includes new contributions added to \cite{nonsusyz}
such that (1) both $N_3$ and $\eta_0$ are studied as dark matter
candidates and (2) the constraints due to neutrino oscillation data are
taken into account in a more extended way than that in \cite{nonsusyz}.

The remaining parts are organized as follows.
In section 2 we address features of the model and discuss a parameter
space consistent with neutrino oscillation data. 
In section 3 we study the relation between the leptogenesis and the CDM
abundance in the model. We examine two possible CDM candidates
taking account of the neutrino oscillation data and the conditions
required by the leptogenesis.  
We will find that the model can give a unified picture 
for the explanation of the neutrino masses, the CDM abundance, and 
the baryon number asymmetry. 
In section 4 we summarize the paper with comments on the signatures 
of the model expected at LHC.

\section{A model for neutrino masses}
The present study is based on the model proposed in \cite{nonsusyz}.
Ingredients of the model and U(1)$^\prime$ charge assignments for these
are given in Table 1. We suppose that U(1)$^\prime$ is
leptophobic.\footnote{We need to introduce some fields to cancel the gauge
anomalies. However, it can be done without affecting the following
study. We present such an example in the Appendix.} 
The extension to general U(1)$^\prime$ is straightforward. 
The fermions listed in Table~1 are assumed to be left-handed.
We note that three singlet fermions $N_{1,2,3}$ are necessary for
present purposes.
Although only two of them are ordered to generate appropriate masses 
and mixing in the neutrino sector, an additional one is necessary 
for the leptogenesis.
The invariant Lagrangian relevant to the neutrino masses can be expressed by 
\begin{eqnarray}
{\cal L}_m&=&\sum_{\alpha=e,\mu,\tau}
\left(h_{\alpha 1}L_\alpha H\bar N_1+h_{\alpha 2}L_\alpha H\bar N_2
+h_{\alpha 3}L_\alpha\eta \bar N_3\right) \nonumber \\
&+&{1\over 2}M_1\bar N_1^2 + {1\over 2}M_2\bar N_2^2 
+{1\over 2}\lambda\phi\bar N_3^2 + {\rm h.c.}. 
\label{masslag}
\end{eqnarray}
Yukawa couplings for charged leptons are assumed to be diagonalized already.
The most general scalar potential invariant under
SU(2)$\times$U(1)$\times$U(1)$^\prime$ gauge symmetry 
up to dimension five is given as
\begin{eqnarray}
V&=&{1\over 2}\lambda_1(H^\dagger H)^2 +{1\over 2}
\lambda_2(\eta^\dagger\eta)^2
+{1\over 2}\lambda_3(\phi^\dagger\phi)^2 \nonumber\\
&+&\lambda_4(H^\dagger H)(\eta^\dagger\eta)
+\lambda_5(H^\dagger\eta)(\eta^\dagger H)
+{1\over 2M_\ast}\left[\lambda_6\phi(\eta^\dagger H)^2 +{\rm h.c.} 
\right]  \nonumber \\
&+& (m_H^2+\lambda_7\phi^\dagger\phi) H^\dagger H 
+(m_\eta^2+\lambda_8\phi^\dagger\phi)\eta^\dagger\eta 
+m_\phi^2\phi^\dagger\phi,
\label{pot}
\end{eqnarray}
where the couplings $\lambda_i$ are real except for $\lambda_6$.
The phase of $\lambda_6$ can induce a physical one which is found 
to be a Majorana phase in the neutrino mass matrix.
A nonrenormalizable $\lambda_6$ term and bare mass terms for $N_{1,2}$ are
added, which will be shown to play crucial roles in the present scenario.
They are supposed to be effective terms generated through some
dynamics at intermediate scales. We assume that $M_\ast\simeq M_1 \ll M_2$
and only $N_1$ and
$N_3$ are related to light neutrino masses and mixings. 

\begin{figure}[t]
\begin{center}
\begin{tabular}{c|cccccccccc}
& $Q_\alpha$ & $\bar U_\alpha$ & $\bar D_\alpha$ & $L_\alpha$ 
& $\bar E_\alpha$ & $\bar N_{1,2}$ & $\bar N_3$ & $H$ & $\eta$ & 
$\phi$ \\\hline
U(1)$^\prime$ & $2q$ & $-2q$ & $-2q$ & 0 & 0 & 0 & $q$ & 0 &  $-q$ 
& $-2q$ \\   \hline 
$Z_2$ & +1 & +1 & +1 & +1 & +1 & +1 & $-1$ & +1 & $-1$ & +1 \\
\end{tabular}
\vspace*{3mm}

{\footnotesize {\bf Table 1.}~~ Field contents and their charges.
$Z_2$ is the residual symmetry of U(1)$^\prime$.}
\end{center}
\end{figure}

The model includes two SU(2) doublet scalars $H$ and $\eta$.
$H$ plays the role of the ordinary doublet Higgs scalar in the SM 
but $\eta$ is assumed to obtain no VEV.
A singlet scalar $\phi$ is also assumed to have a real VEV at suitable
scales, 
which breaks U(1)$^\prime$ down to $Z_2$. 
The $Z_2$ charge for each field can be found in Table 1.
The VEV of $\phi$ gives masses for $N_3$ and $Z^\prime$ as
\begin{equation}
M_{N_3}=\lambda\langle\phi\rangle, \qquad 
M_{Z^\prime}=2\sqrt{2}g^\prime q\langle\phi\rangle, 
\label{nz}
\end{equation}
where $\lambda$ is assumed to be real. Since $M_{Z^\prime}$ is bounded
from below by the $Z^\prime$ phenomenology, $M_{N_3}$ has also lower
bounds for fixed values of $\lambda$.  
It also yields an effective coupling constant 
$\lambda_6\langle\phi\rangle/M_\ast$ in the $\lambda_6$ term.
It can be small enough to make radiative neutrino masses tiny 
even for $O(1)$ values of $\lambda_6$ 
as long as $\langle\phi\rangle\ll M_\ast$ is satisfied.
Since the mixing between $\eta^0$ and $\eta^{0\ast}$ is induced through
this small coupling, the mass eigenvalues split slightly. 
The states $\chi_\pm^0\equiv{1\over\sqrt 2}
\left(\eta^0 \pm \eta^{0\ast}\right)$
have mass eigenvalues such as  
\begin{eqnarray}
M_{\chi_\pm^0}^2&=& m_\eta^2 +(\lambda_4+\lambda_5)\langle H^0\rangle^2
+\lambda_8\langle\phi\rangle^2 
\pm {|\lambda_6|\langle\phi\rangle\over M_\ast}\langle H^0\rangle^2 
\nonumber\\
&\equiv& M_\eta^2 \pm {|\lambda_6|\langle\phi\rangle\over M_\ast}
\langle H^0\rangle^2.
\label{etamass}
\end{eqnarray}
The magnitude of the difference of these eigenvalues is constrained by
the direct search of the CDM if either of these $\chi_\pm^0$ is 
the lightest $Z_2$ odd field. 
Mass of the charged states $\eta^\pm$ is given by
\begin{equation}
M_{\eta^\pm}=m_\eta^2 + \lambda_4 \langle H^0\rangle^2 
+\lambda_8\langle\phi\rangle^2,
\end{equation}
and then $M_{\chi_\pm^0}$ can be much smaller than $M_{\eta^\pm}$
in case of $\lambda_5<0$. These points will be discussed in the analysis of
the CDM later.
Since $\lambda_6$ is complex in general, the $CP$ violation may be
detected through this $\eta^0$-$\eta^{0\ast}$ mixing.
Although this is an interesting feature of the model, 
we do not discuss this subject further in this paper.

We have two distinct origins for the neutrino masses in this model. 
One is the ordinary seesaw mass induced by a right-handed neutrino 
$N_1$ \cite{sterile}. 
Another one is the one-loop radiative mass mediated by the exchange 
of $\eta^0$ and $N_3$ \cite{z2,radmass}.
Although $N_2$ also has contributions to the neutrino mass generation
through the seesaw mechanism, its effect can be safely neglected
compared with these if $M_2$ is large enough.
However, baryogenesis caused by leptogenesis requires this contribution
since $N_3$ is has no lepton number as discussed below.
The radiative neutrino mass generation requires some lepton number
violation.
We can put them either in ${\cal L}_m$ or $V$.
If we assume that $\eta$ and $N_3$ have the lepton number $-1$ and $0$, 
respectively, the $\lambda_6$ term in $V$ brings about this required lepton 
number violating effect. We adopt this choice in the following arguments.
$N_{1,2}$ are considered to have lepton number +1.
   
The mass matrix for three light neutrinos induced by these origins  
is summarized as
\begin{equation}
M_\nu={\langle H^0\rangle^2\over M_\ast}\left[\mu^{(1)}
+{\lambda_6\over 8\pi^2\lambda}
I\left({M_{N_3}^2\over M_{\eta_0}^2}\right)\mu^{(3)}\right], \qquad
I(t)={t\over 1-t}\left(1+{t\ln t\over 1-t}\right),
\label{nmass}
\end{equation}
where $\mu^{(a)}$ is defined by
\begin{equation}
\mu^{(a)}=\left(\begin{array}{ccc}
h_{ea}^2 & h_{ea}h_{\mu a} & h_{ea}h_{\tau a} \\
h_{ea}h_{\mu a} & h_{\mu a}^2 & h_{\mu a}h_{\tau a} \\
h_{ea}h_{\tau a} & h_{\mu a}h_{\tau a} & h_{\tau a}^2 \\
\end{array}\right) \quad (a=1,3).
\label{yukawa}
\end{equation}
Both $h_{\alpha 1}$ and $h_{\alpha 3}$ are assumed to be real, for simplicity.
We note that two terms in $M_\nu$ have the similar texture although they 
are characterized by different mass scales. 
If we impose commutativity between $\mu^{(1)}$ and $\mu^{(3)}$, 
the condition 
\begin{equation}
h_{e1}h_{e3}+h_{\mu 1}h_{\mu 3}+h_{\tau 1}h_{\tau 3}=0
\label{cyukawa}
\end{equation}
is needed to be satisfied. 
We consider this simple case in the following as an interesting
example, since it allows us to study
the mass matrix analytically.\footnote{If nonzero eigenvalues are
dominated by different origins respectively, this will be a good
approximation to describe such cases.}

We introduce a matrix $\tilde U$ to diagonalize the larger term
of $M_\nu$ at first, which is defined as 
\begin{equation}
\tilde{U}=
\left(\begin{array}{ccc}
1 & 0 & 0 \\
0 & \cos\theta_{2} &\sin\theta_{2} \\
0 &-\sin\theta_{2} &\cos\theta_{2} \\
\end{array}\right)
\left(\begin{array}{ccc}
\cos\theta_{3} & 0 &\sin\theta_{3} \\
0 & 1 & 0 \\
-\sin\theta_{1} & 0 & \cos\theta_{3}\\
\end{array}\right).
\end{equation}
Then the matrix $\mu^{(a)}$ in $M_\nu$ can be diagonalized as 
$\tilde U^T \mu^{(a)}\tilde U$ if the angles $\theta_{2,3}$ satisfy 
\begin{equation}
\tan\theta_2={h_{\mu a}\over h_{\tau a}}, \qquad 
\tan\theta_{3}={h_{ea}\over \sqrt{h_{\mu a}^2+h_{\tau a}^2}}.
\label{angle}
\end{equation}
Eigenvalues for this matrix are found to be
\begin{equation}
\mu_{\rm diag}^{(a)}={\rm diag}(0,~0,~h_{ea}^2
+h_{\mu a}^2+h_{\tau a}^2).
\label{first}
\end{equation}
Another term $\mu^{(a^\prime)}$ is also transformed by 
$\tilde U$. However, 
if the condition (\ref{cyukawa}) is satisfied,
$\mu^{(a^\prime)}$ can be diagonalized by an orthogonal transformation 
$\tilde UU_1$ supplemented by an additional transformation
\begin{equation}
U_1=\left(\begin{array}{ccc}
\cos\theta_1 &\sin\theta_1 & 0  \\
-\sin\theta_1 &\cos\theta_1 & 0 \\
0 & 0 & 1 \end{array}\right),
\end{equation}
and we have eigenvalues
\begin{equation}
\mu_{\rm diag}^{(a^\prime)}={\rm diag}(0,~h_{ea^\prime}^2
+h_{\mu a^\prime}^2+h_{\tau a^\prime}^2,~0).
\end{equation}
Here $\theta_1$ is defined as
\begin{equation}
\tan\theta_1=-{\tan\tilde\theta_2\tan\theta_2+1\over
(\tan\tilde\theta_2-\tan\theta_2)\sin\theta_3}, \qquad
\tan\tilde\theta_2={h_{\mu a^\prime}\over h_{\tau a^\prime}}.
\label{ue3}
\end{equation}
We note that this $U_1$ transformation does not affect the diagonalization 
of $\mu^{(a)}$. 
 
If we define the mass eigenvalues as 
$ U^T M_\nu U= {\rm diag}(0, m_2,m_3)$ where $m_2<m_3$ is assumed,
they can be written as
\begin{equation}
m_2=AB~{\tan^2\theta_1+1\over \tan^2\theta_2+1}
(\tan\tilde\theta_2-\tan\theta_2)^2, \quad
m_3={A\over 2}(\tan^2\theta_2+1)(\tan^2\theta_3+1). 
\label{mass1}
\end{equation}
Here we find that there are two possibilities for generation of
$m_3$ and $m_2$.
The first case is realized by taking $a=1$ and $a^\prime=3$ in the above
formulas, and then $m_3$ is induced by the ordinary 
seesaw mechanism. In this case
$A$ and $B$ are defined by  
\begin{equation}
{\rm (i)}\qquad A\equiv {2h^2_{\tau 1}\langle H^0\rangle^2\over M_\ast}, 
\qquad 
B\equiv {|\lambda_6|\over 8\pi^2\lambda}
\left({h_{\tau 3}\over h_{\tau 1}}\right)^2
I\left({M_{N_3}^2\over M_{\eta_0}^2}\right).
\end{equation}
The second case is obtained by taking $a=3$ and $a^\prime=1$, and then
$m_3$ is determined by the radiative effect.
In this case $A$ and $B$ are written as
\begin{equation}
{\rm (ii)}\qquad A\equiv {h^2_{\tau 3}\langle H^0\rangle^2\over M_\ast}
{|\lambda_6|\over 4\pi^2\lambda}I\left({M_{N_3}^2\over M_{\eta_0}^2}\right), 
\qquad 
B\equiv \left[{|\lambda_6|\over 4\pi^2\lambda}
I\left({M_{N_3}^2\over M_{\eta_0}^2}\right)\right]^{-1}
\left({h_{\tau 1}\over h_{\tau 3}}\right)^2.
\end{equation}

\input epsf
\begin{figure}[t]
\begin{center}
\epsfxsize=7cm
\leavevmode
\epsfbox{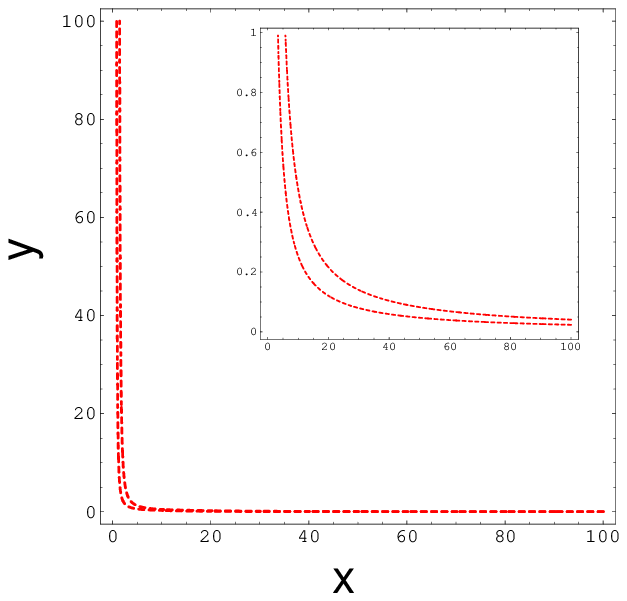}
\hspace*{7mm}
\epsfxsize=7cm
\leavevmode
\epsfbox{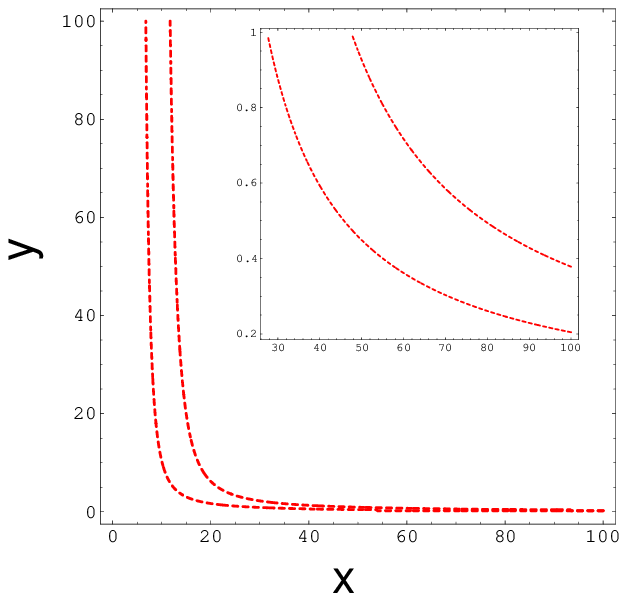}
\vspace*{-2mm}
\end{center}
{\footnotesize {\bf Fig.~1}~~The region in the $(x, y)$ plane 
allowed by the neutrino oscillation data. 
The cases (i) and (ii) correspond to the left-handed and right-handed 
panel, respectively. The figure focused to the $0<y<1$  region is 
also displayed in each panel.}
\end{figure}

Since only two mass eigenvalues can be considered nonzero in the present
setting, 
neutrino oscillation data require that these mass eigenvalues should satisfy
$m_3=\sqrt{\Delta m_{\rm atm}^2}$ and 
$m_2=\sqrt{\Delta m_{\rm sol}^2}$ \cite{nmass}.
Data of the atmospheric neutrino and the K2K experiment require 
$\tan\theta_2=1$. We also find that $\theta_1$ should be taken as 
$\theta_{\rm sol}$ which is a
mixing angle relevant to the solar neutrino. 
The CHOOZ experiment gives a constraint on $\theta_3$ 
such as $|\sin\theta_3|<0.22$ \cite{chooz}. 
If we use these conditions, the mixing matrix $U=\tilde{U}U_1$ 
can be approximately written as 
\begin{equation}
U=\left(\begin{array}{ccc}
\cos\theta_{\rm sol} &\sin\theta_{\rm sol} & {\sin\theta_3\over\sqrt{2}}  \\
-{\sin\theta_{\rm sol}\over\sqrt{2}} &{\cos\theta_{\rm sol}\over\sqrt{2}} & 
{1\over\sqrt{2}} \\
{\sin\theta_{\rm sol}\over\sqrt{2}} &-{\cos\theta_{\rm sol}\over\sqrt{2}} & 
{1\over\sqrt{2}} \\
\end{array}\right).
\end{equation}
By imposing the experimental values on
$\tan\theta_{\rm sol}$, $\sqrt{\Delta m_{\rm atm}^2}$,
$\sqrt{\Delta m_{\rm sol}^2}$, and $\sin\theta_3$, we can constrain the
values of $A$ and $B$ \cite{nonsusyz}.
For simplicity, we assume $\lambda=|\lambda_6|$. 

The condition for $A$ constrains the Yukawa coupling $h_{\tau 1}$ as
\begin{eqnarray}
&&{\rm (i)}\quad h_{\tau 1}\simeq 2.9\times 10^{-4}
\left({M_\ast  \over 10^{8}{\rm GeV}}\right)^{1/2}, \nonumber \\
&&{\rm (ii)}\quad 7.9 \times 10^{-5}
\left({M_\ast  \over 10^{8}{\rm GeV}}\right)^{1/2}
~{^<_\sim}~h_{\tau 1}~{^<_\sim}~1.3 \times 
10^{-4}\left({M_\ast \over 10^{8}{\rm GeV}}\right)^{1/2}. 
\label{tau1}
\end{eqnarray}
If we require $h_{\tau 1}$ and $h_{\tau 3}$ to
be in perturbative regions, we find that both $M_\ast$ and $M_\ast x^2$
should be less than $10^{16}$~GeV.
Here we introduce two parameters $x\equiv h_{\tau 3}/ h_{\tau 1}$ 
and $y\equiv M_{N_3}/M_{\eta}$. 
The condition for $B$ selects the regions in the $(x,y)$ plane which are 
consistent with the neutrino oscillation data. 
They are shown for both cases (i) and (ii) as the regions sandwiched 
by the dashed lines in Fig.~1.
These figures show that the model can explain the neutrino oscillation
data in rather wide parameter regions.
In particular, it is useful to note in relation to the CDM that
we can have solutions for large values of $y$ such as $10^6$ as long as 
$x$ stays in the constrained region: (i)~$0.55-0.8$ and (ii) $3.5-6.5$. 
By using these results obtained from the neutrino oscillation 
data, we examine the leptogenesis and the CDM abundance 
in this model in the next section.
 
\section{Leptogenesis and CDM abundance}
The present model contains several new neutral fields 
with nonzero lepton number or 
an odd $Z_2$ charge. Thus, we have sufficient ingredients with the
required properties for both 
leptogenesis and CDM candidates. 
Although one might consider that there are several scenarios for these 
explanations in this model,
they seem to be constrained by the neutrino oscillation data.

The lightest neutral field with an odd $Z_2$ charge can 
be stable and then a CDM candidate since an even charge is assigned 
to each SM content. 
If $y<1$ is satisfied,  $N_3$ can be a CDM candidate.
As in the ordinary leptogenesis scenario, $N_1$ related to the ordinary
seesaw mechanism can be a mother field for leptogenesis. 
However, since two right-handed neutrinos 
are necessary to realize the $CP$ asymmetry, we need to introduce 
$N_2$ with the lepton number $+1$ as mentioned before. 

On the other hand, since $\eta^0$ has both the odd $Z_2$ charge and 
the lepton number, it
might be considered as the origin of the CDM or the lepton number
asymmetry in the case of $y>1$.
However, it might be difficult to contribute both of them since 
it has the SM gauge interactions.
The situation is similar to sneutrinos in the supersymmetric models. 
Sneutrinos have been rejected to be a CDM candidate through 
the direct detection experiments. 
This constraint might be escapable in the $\eta^0$ case since there is
the $\eta^0$-$\eta^{0\ast}$ mixing due to the $\lambda_6$ term which
generates the mass difference between its components.
The model has to satisfy suitable conditions for this mass difference 
if this possibility is realized.
On the other hand, this $\eta^0$ is too light to be a mother field for
sufficient production of the lepton number asymmetry 
through the out-of-equilibrium decay,
although the $\eta^0$ sector can bring the almost 
degenerate mass eigenstates through the $CP$ violating mixing and cause
the resonant decay. We examine these subjects in detail below.

\subsection{Leptogenesis}
If we take account of the existence of $N_2$ which can be neglected in
the estimation of the neutrino masses,
the leptogenesis is expected to occur through the decay of $N_1$.
In fact, it is heavy enough for the out-of-equilibrium decay and it has the 
lepton number violation through a Majorana mass term.
By taking account of the well known relation $B=28(B-L)/75$ which comes
from re-processing of the $B-L$ asymmetry by sphaleron transitions, 
the generated baryon number asymmetry is given by
\begin{equation}
{n_B\over s}=-{28\over 75}Y^{\rm eq}_{N_1}\varepsilon\kappa, 
\end{equation}
where $Y^{\rm eq}_{N_1}(\equiv n_{N_1}/s)$ is the ratio of the equilibrium 
number density of $N_1$ to the entropy density. 
The $CP$ asymmetry in the $N_1$ decay and 
the wash-out effect are represented by $\varepsilon$ and
$\kappa$, respectively. 
If temperature is much larger than $M_1$, we have
$Y^{\rm eq}_{N_1}\simeq 0.42/g_\ast$ by using
$n_{N_1}=(3\zeta(3)/2\pi^2)T^3$ and $s=(2\pi^2g_\ast/45)T^3$.
The relativistic degrees of freedom in this model is 
$g_\ast\simeq 130$. Thus, the $CP$ asymmetry $\varepsilon$ required to
produce the present baryon number asymmetry is estimated as 
\begin{equation}
\varepsilon\simeq -7.2\times 10^{-8}\kappa^{-1}, 
\label{leptg}
\end{equation}
where we use  $n_B/s\simeq (0.87\pm 0.04)\times 10^{-10}$ which 
is predicted by nucleosynthesis and CMB measurements \cite{basym}. 
The $CP$ violation in the $N_1$ decay is induced through 
interference between the tree and one-loop amplitudes.
This induced $CP$ asymmetry $\varepsilon$ is estimated as \cite{leptg2}
\begin{equation}
\varepsilon=-{3\over16\pi}{M_1\over M_2}
{{\rm Im }[(h^\dagger h)_{12}^2]\over |h^\dagger h|_{11}}.
\end{equation}

Now we estimate $\varepsilon$ in this model.
As discussed in the previous section, there are two ways for
generation of the neutrino masses $m_3$ and $m_2$. 
The $CP$ asymmetry $\varepsilon$ 
can also have different values for these two cases.
For simplicity, we assume $|h_{\alpha 2}|\simeq |h_{\alpha 1}|$.
This does not affect the estimation of the neutrino masses because of
the assumed setting $M_\ast\simeq M_1 \ll M_2$. 
In that case we have
\begin{equation}
\left|{\rm Im}[(h^\dagger h)_{12}^2]\right|~{^<_\sim}~4 h_{\tau 1}^4~\simeq
\left\{
\begin{array}{ll}
\displaystyle
2.8\times 10^{-14}\left({M_\ast\over 10^{8}{\rm GeV}}\right)^2 & 
{\rm for ~(i)}, \\
\displaystyle
(0.16-1.1)\times 10^{-14}\left({M_\ast\over 10^{8}{\rm GeV}}\right)^2 & 
{\rm for~(ii)},
\end{array}\right.
\end{equation}
where we apply the results in eq.~(\ref{tau1}) to this estimation.
We use these maximum values for ${\rm Im}[(h^\dagger h)_{12}^2]$ 
in the formulas of $\varepsilon$ here.

In case (i), we have the relation
$|h^\dagger h|_{11}\langle H_0\rangle^2/
M_\ast\simeq\sqrt{\Delta m_{\rm atm}^2}$  
and then $\varepsilon$ can be written as
\begin{equation}
\varepsilon\simeq -9.8 \times 10^{-8}
\left({10^{10}\kappa^{-1}{\rm GeV}\over M_2}\right) 
\left({M_\ast\over 10^{8}~{\rm GeV}}\right)^2\kappa^{-1}.
\label{asym}
\end{equation}
In case (ii), we note that the seesaw mechanism gives $m_2$ and
the relation 
$|h^\dagger h|_{11}\langle H_0\rangle^2/
M_\ast\simeq\sqrt{\Delta m_{\rm sol}^2}$
is satisfied. Thus, we find that $\varepsilon$ is expressed as
\begin{equation}
\varepsilon= -2.2 \times 10^{-8}
\left({10^{10}\kappa^{-1}{\rm GeV}\over M_2}\right) 
\left({M_\ast\over 10^{8}~{\rm GeV}}\right)^2\kappa^{-1}.
\end{equation}
These results show that a sufficient $CP$ asymmetry can be generated
for 
\begin{equation}
M_\ast\simeq \left\{
\begin{array}{ll}
\displaystyle 8.6\times 10^{7}\left({M_2 \over 10^{10}\kappa^{-1}~{\rm GeV}}
\right)^{1/2}~{\rm GeV} &\quad {\rm for~(i)}, \\
\displaystyle 1.8\times 10^{8}\left({M_2 \over 10^{10}\kappa^{-1}~{\rm GeV}}
\right)^{1/2}~{\rm GeV} &\quad  {\rm for~(ii)}.
\end{array}\right.
\end{equation}
Consistency with the present setting $M_2 \gg M_\ast$ can be satisfied 
for $M_2~{^>_\sim}~ 10^{10}\kappa^{-1}$~GeV in both cases, for example. 
It may be useful to remind that $\kappa$ is expected to be 
$10^{-1}-10^{-3}$ from the numerical study of the Boltzmann equation.
Such an analysis also shows that the leptogenesis is possible only 
for narrow ranges of 
$\tilde m_1=|h^\dagger h|_{11}\langle H_0\rangle^2/M_1$ \cite{leptg2}.
In the present model this $\tilde m_1$ is estimated as
\begin{eqnarray}
\tilde m_1\simeq\left\{\begin{array}{ll}
\displaystyle \sqrt{\Delta m_{\rm atm}^2}~{M_\ast\over M_1} 
&\quad  {\rm for~~(i)}, \\
\displaystyle \sqrt{\Delta m_{\rm sol}^2}~{M_\ast\over M_1} 
&\quad  {\rm for~~(ii)}.
\end{array}\right.
\end{eqnarray}
This suggests that $M_\ast~{^<_\sim}~M_1$ is favored by
leptogenesis and it could be consistent in the present settings. 
The values of $M_\ast/M_1$ determine which case between them is more promising.
These results show that the out-of-equilibrium decay of $N_1$ can
produce the necessary baryon number asymmetry for intermediate values 
of $M_1$ as in the usual cases. 
As long as we confine ourselves to the non-supersymmetric framework, 
the model is free from the gravitino problem. 

\subsection{CDM candidates and their abundance}
The lightest field with an odd $Z_2$ charge can be stable since the even
charge is assigned to each SM content. If both the mass
and the annihilation cross section of such a field have appropriate values, 
it can be a good CDM candidate as long as it is neutral.
As mentioned before, we have two such candidates, that is, the lighter
one of $\chi_\pm^0$ (we represent it by $\chi_L^0$) and $N_3$. 

At first, we consider the $y<1$ case in which $N_3$ is the CDM. 
Its annihilation is expected to be mediated by both the exchange of 
$\eta^0$ and the U(1)$^\prime$ gauge boson.
If their annihilation is mediated only by the former one 
through Yukawa couplings as in the model discussed in \cite{nonsusy2},
we need fine tuning of coupling constants to explain both the observed 
value of the CDM abundance and the constraints coming from lepton 
flavor violating processes such as $\mu\rightarrow e \gamma$.
However, in the present case the $N_3$ annihilation can be dominantly 
mediated by the U(1)$^\prime$ gauge interaction since 
Yukawa coupling constants $h_{\alpha 3}$ can be small enough as
estimated in eq.~(\ref{tau1}). 
Thus, we may expect that $N_3$ can cause the satisfactory 
relic abundance as the CDM in rather wide parameter regions .
We also note that the U(1)$^\prime$ is supposed to be a generation 
independent gauge symmetry and then the FCNC problem can be easily 
escaped in this case. 

In order to estimate the $N_3$ abundance, we consider to expand 
the annihilation cross section for $N_3N_3 \rightarrow f\bar f$ by the 
relative velocity $v$ between the annihilating $N_3$ 
as $\sigma v= a + b v^2$. 
The coefficients $a$ and $b$ are expressed as
\begin{equation}
a=\sum_fc_f{g^{\prime 4}\over 2\pi}Q_{f_A}^2 q^2
{m_f^2\beta\over (s-M_{Z^\prime}^2)^2}, \qquad
b=\sum_fc_f{g^{\prime 4}\over 6\pi}(Q_{f_V}^2+Q_{f_A}^2)
q^2{M_{N_3}^2\beta \over (s-M_{Z^\prime}^2)^2},
\end{equation}
where $\beta=\sqrt{1-m_f^2/M_{N_3}^2}$ and $c_f$=3 for quarks. 
$s$ is the center of mass energy of collisions and $q$ is the U(1)$^\prime$ 
charge of $N_3$ given in Table 1. 
The charge of the final state fermion $f$ is defined as
\begin{equation}
Q_{f_V}=Q_{f_R}+Q_{f_L}, \qquad Q_{f_A}=Q_{f_R}-Q_{f_L}.
\end{equation}
Using these quantities, the present relic abundance of $N_3$ can be 
estimated as \cite{cdm},
\begin{equation}
\Omega_{N_3} h^2|_0=
\left.{M_{N_3} 
n_{N_3}\over \rho_{\rm cr}/h^2 }\right|_0
\simeq{8.76\times 10^{-11}g_\ast^{-1/2}x_F\over 
(a+3b/x_F)~{\rm GeV}^2 }.
\end{equation} 
where $g_\ast$ is the degrees of freedom of relativistic 
fields at the freeze-out temperature $T_F$ of $N_3$.
The dimensionless parameter $x_F=M_{N_3}/T_F$ is determined through the
condition 
\begin{equation}
x_F=\ln{0.0955m_{\rm pl}M_{N_3}(a+6b/x_F)
\over (g_\ast x_F)^{1/2}},
\end{equation}  
where $m_{\rm pl}$ is the Planck mass. 
If we fix the U(1)$^\prime$ charge of the relevant fields and  
its coupling constant $g^\prime$, we can estimate 
the present $N_3$ abundance using these formulas. It can be
compared with $\Omega_{N_3}h^2=0.1045^{+0.0072}_{-0.0095}$
given by the three year WMAP \cite{wmap3}. 

\begin{figure}[t]
\begin{center}
\epsfxsize=7.5cm
\leavevmode
\epsfbox{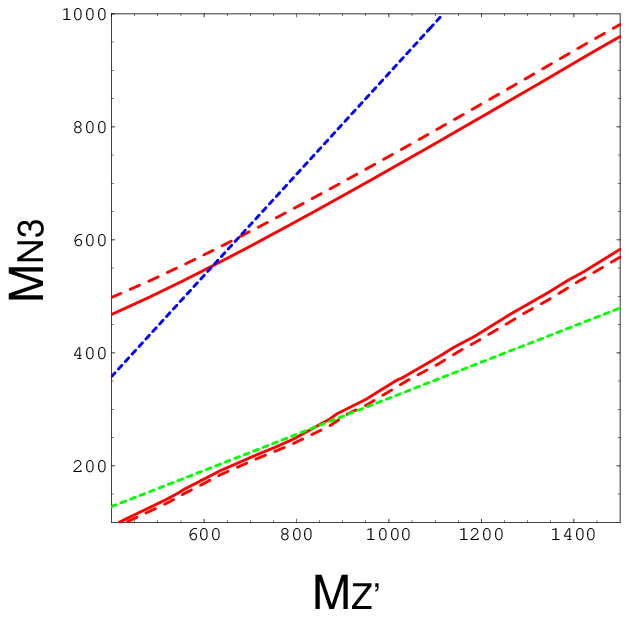}
\vspace*{-2mm}
\end{center}
{\footnotesize {\bf Fig.~2}~~Regions allowed by the WMAP data in the
 $(M_{Z^\prime}, M_{N_3})$ plane. Green and blue dotted lines 
represent $M_{N_3}$ lines for $\lambda=0.25$ and 0.7, respectively.}
\end{figure}

We numerically examine the possibility that the CDM abundance 
is consistently explained in this model. 
We use the GUT relation $g^\prime=\sqrt{5/3}g_1$ and $q=0.6$ as an example.
The regions in the $(M_{Z^\prime}, M_{N_3})$ plane allowed by the WMAP
data are shown in Fig.~2. 
They appear as two narrow bands sandwiched by both a solid line 
and a dashed line. 
The lower bounds of $M_{Z^\prime}$ come from constraints for $ZZ^\prime$
mixing and a direct search of $Z^\prime$.
Since the Higgs field $H$ is assumed to have no U(1)$^\prime$ charge,
its VEV induces no $ZZ^\prime$ mixing. Moreover, since it is assumed to
be leptophobic, the constraint on $M_{Z^\prime}$ obtained from 
its hadronic decay is rather weak. The lower bounds of 
$M_{Z^\prime}$ may be $M_{Z^\prime}~{^>_\sim}~450$~GeV 
in the present model \cite{leptph}. 
Since the masses of $Z^\prime$ and $N_3$ are correlated through
eq.~(\ref{nz}),
we can draw a line of $M_{N_3}$ in the $(M_{Z^\prime}, M_{N_3})$ 
plane by fixing a value of $\lambda$.
In Fig.~2, such lines are represented by the green and blue 
dotted ones for $\lambda=0.25$ and 0.7, respectively. 
For these $M_{N_3}$ values required by the WMAP, $M_\eta$ is found to 
take values such as $\sim 300/y$ GeV and $\sim 580/y$ GeV for 
$\lambda=0.25$ and 0.7. 
Using Figs.~1 and 2, we can determine the range of $x$, 
if $M_\eta$ and then $y$ is fixed. We find 
that $x$ takes very restricted values for the case of 
$M_\eta~{^<_\sim}~1$~TeV, especially in case (i). 

In Fig~2 we can observe an interesting feature of $Z^\prime$. 
Although we assume it is leptophobic, it can have nonhadronic decay model
as long as $2M_{N_3}<M_{Z^\prime}$ is satisfied. Fig.~2 shows that this
condition is satisfied only at the lower allowed band
 but not at the upper allowed band. 
Thus, $Z^\prime$ can have nonhadronic decay mode only for
$\lambda~{^<_\sim}~0.33$. 

\begin{figure}[t]
\begin{center}
\epsfxsize=7.5cm
\leavevmode
\epsfbox{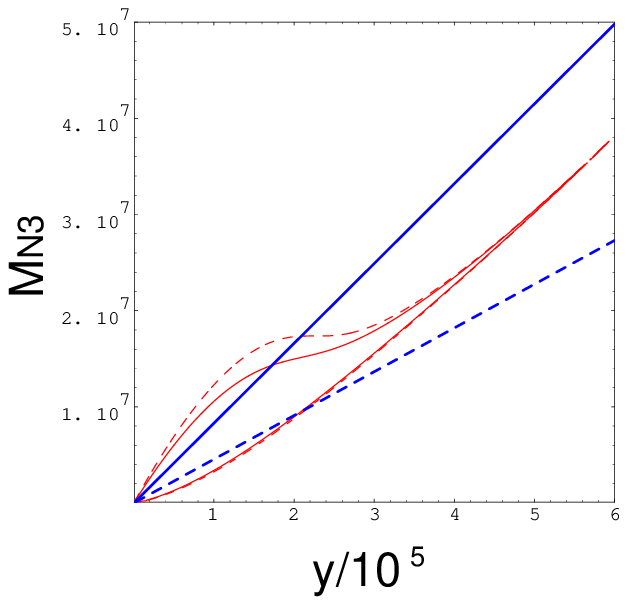}
\vspace*{-2mm}
\end{center}
{\footnotesize {\bf Fig.~3}~~Allowed regions in the
 $(y, M_{N_3})$ plane. A red thin dotted line and a red thin solid line 
corresponds to an upper and lower bound of $\Omega_{\chi_L^0}h^2$ 
imposed by the WMAP data. A blue thick solid line represents a line for 
$M_{\chi_L^0}=80$~GeV. A blue thick dotted line represents a boundary for
$M_{\chi_+^0}+M_{\chi_-^0}=m_Z$.}
\end{figure}

If $y>1$ is satisfied, the neutral scalar $\chi_L^0$ is the CDM.
In this case we can follow the analysis given in \cite{bar}.
If it is heavier than the $W^\pm$ boson, it cannot keep the 
relic abundance required from the WMAP data.
The reason is that they can effectively annihilate to the $W^\pm$ 
pair through the $Z^0$ exchange. 
Thus, since we have no other candidate for the CDM within the present model,
we have to assume that the mass of $\chi_L^0$ should be smaller
than 80 GeV. 
Even if it is lighter than the $W^\pm$ boson, direct search
experiments impose a strong constraint.
The difference of the mass eigenvalues of $\chi_\pm^0$ is estimated as
\begin{equation}
\Delta M\simeq {|\lambda_6|\langle\phi\rangle\over M_\eta M_\ast}
\langle H^0\rangle^2 
\sim {M_{N_3}\over M_\eta M_\ast}\langle H^0\rangle^2
\sim 300 y 
\left({10^{8}~{\rm GeV}\over M_\ast}\right)~{\rm keV}.
\label{mdel}
\end{equation}
Since the $\chi_\pm^0$ have a vector like interaction with $Z^0$ boson,
its elastic scattering cross section with a nucleon through $Z^0$ exchange
is 8-9 orders of magnitude larger than the existing direct search 
limits \cite{direct}.
To forbid $Z^0$ exchange kinematically, $\Delta M$ has to be larger than
a few 100 keV \cite{zexchange}. 
Following eq.~(\ref{mdel}), this constraint can be interpreted as
a condition $y~{^>_\sim}~(M_\ast/10^{8}~{\rm GeV})$.

If we impose that the relic $\chi_L^0$ abundance saturates the values required
by the WMAP data, a much stronger constraint can be obtained.
This $\chi_L^0$ abundance is dominantly determined by the $p$-wave
suppressed coannihilation process $\chi_+^0\chi_-^0
\rightarrow Z^\ast \rightarrow \bar f f$. In order to realize a 
suitable relic abundance, we need to decrease this coannihilation rate
by requiring the heavier one of $\chi_\pm^0$ is thermally suppressed. 
This requires that $\Delta M~{^>_\sim}~8-9$~GeV should be satisfied for
$M_{\chi_L^0}=60-73$ GeV \cite{bar}.
Thus, if we consider  $\chi_L^0$ is the CDM taking account of this arguments,
we have an another condition $y~{^>_\sim}~ M_\ast/(3000~{\rm GeV})$.
Since the leptogenesis occurs successfully for $M_\ast~{^>_\sim}~10^9$~GeV as
seen in the previous part, $y$ should be a larger value than
$2\times 10^5$ and then $M_{N_3}$ should be larger than $3\times 10^7$~GeV.

We can search favored parameter regions in the present model 
by estimating numerically the relic abundance of $\chi_L^0$ 
in the same way as the $N_3$ case. 
In this estimation we need to take account of the above
mentioned thermal effect which modifies the relic density in the 
$\Delta M=0$ case by a factor ${1\over 2}\exp(\Delta M/T_F)$.
In Fig.~3 we plot the allowed regions in the $(y, M_{N_3})$ plane for 
the case of $M_\ast=10^9$~GeV, which is a favored value for leptogenesis.
In the regions sandwiched by both dotted and solid thin lines, 
$\Omega_{\chi_L^0}$ realizes the three year WMAP data.
In the same figure we add two conditions.
We plot a line corresponding to $M_{\chi_L^0}=80$~GeV by a blue 
solid thick one.
Since we now consider regions below the $WW$ threshold, 
allowed regions are the part below this line. 
The $Z^0$ width also imposes an another
condition $M_{\chi_+^0}+M_{\chi_-^0}>m_Z$. The boundary of this
condition is plotted by a blue dotted thick line. Regions
above this boundary satisfy this condition.  
As seen from this figure, the favored part in the regions sandwiched 
by these thick lines gives $40-80$~ GeV for $M_{\chi_L^0}$, 
which agrees with the results given in 
\cite{bar,zexchange}. This does not contradict with
experimental mass bounds for charged Higgs fields as long as $\lambda_4$ has
suitable negative values.  
The constraint from $\mu\rightarrow e\gamma$ can be also satisfied for
$M_\ast$ which can keep Yukawa couplings small enough in eq.~(\ref{tau1}).
For the required large values $(2-5)\times 10^5$ for $y$, 
$|\lambda_6|\langle\phi\rangle\ll M_\ast$ can be still satisfied and  
$Z^\prime$ becomes very heavy so as to be out of the range 
reached by the LHC experiments. \footnote{In the original models
\cite{nonsusy2}, required values of $\Delta M$ and $M_{\chi_L^0}$
for the $\chi_L^0$ CDM can be consistent with 
the neutrino oscillation data and the FCNC constraint as long as singlet
fermion masses are large enough and their Yukawa couplings are small as
in the present case. Thus, we could not find substantial difference
between this model and the original ones in the $y>1$ case.}
In this case $x$ is confined to very restricted regions, 
especially in case (i).
In order to realize the favorable values of $M_{\chi_L^0}$ and $\Delta
M$, several coupling constants are required to be finely tuned. For example,
$\lambda_8$ should be very small like $O(10^{-5})$.
Although these required parameter tuning might decrease interests 
for this case compared with the $y<1$ case,
it is noticeable that $\chi_L^0$ can be a CDM candidate consistently with
the neutrino oscillation data in this model.
  
\section{Summary}
We have studied a unified explanation for both the CDM abundance and the baryon
number asymmetry in a non-supersymmetric model for neutrino masses. 
The model is obtained from the SM by adding a U(1)$^\prime$ gauge
symmetry and several neutral fields. 
The neutrino masses are generated through both the seesaw mechanism 
with a single right-handed neutrino and the one-loop radiative 
effects. Both contributions induce the same texture which can realize 
favorable mass eigenvalues and mixing angles. 
New neutral fields required for this mass generation make the unified
explanation for the leptogenesis and the CDM abundance in the universe 
possible. 

Both the neutral fermion $N_3$ and the neutral scalar $\eta^0$ 
are stable due to a $Z_2$ subgroup which remains as a 
residual symmetry of the spontaneously broken U(1)$^\prime$.
Thus, they can be a good CDM candidate. 
In the $N_3$ CDM case, since it has the U(1)$^\prime$ gauge interaction, 
the annihilation of this CDM candidate 
is dominantly mediated through this interaction. 
If this U(1)$^\prime$ symmetry is broken at a scale suitable for the
neutrino mass generation, 
its estimated relic abundance can explain the WMAP result 
for the CDM abundance. We examined these points taking account of the
neutrino oscillation data.
In the $\eta^0$ CDM case, if it is lighter than $W^\pm$ boson and the
difference of its mass eigenstates forbid its coannihilation due to 
the $Z^0$ exchange
kinematically, it can keep the suitable relic abundance. We examined the
consistency of this picture with the neutrino oscillation data.

Since another introduced neutral fermion $N_1$ is a gauge
singlet and heavy enough, it can follow the out-of-equilibrium decay 
which produces the baryon number asymmetry through the leptogenesis.  
We showed the consistency of this scenario with the neutrino oscillation data.
Although the required reheating temperature for the leptogenesis 
is similar values to the one in the ordinary seesaw mechanism, 
we have no gravitino problem since we need no supersymmetry to
prepare the stable CDM candidates.
The present model gives an example in which three of the biggest 
experimental questions in the SM, that is, neutrino masses, the CDM abundance, 
and the baryon number asymmetry can be explained through the closely 
related physics in a non-supersymmetric extension of the SM. 
In order to solve the hierarchy problem, a supersymmetric extension 
of the model may be considered along the
line of \cite{proneu}. 
We would like to discuss this subject elsewhere.

Finally, we briefly comment on signatures of the model expected at LHC.
The above study fixes mass spectrum of the relatively light fields 
in the model. We have $N_3$, $\eta$ and $Z^\prime$ as such new fields. 
$\eta$ is expected to be produced through the $W$ fusion 
as in the similar way to the ordinary Higgs field. 
Since $\eta$ has Yukawa couplings with leptons only, its components
$\eta^0$ and $\eta^\pm$ can be distinguished from others such as the 
Higgs fields in the MSSM through the difference of the decay modes. 
$Z^\prime$ couples with quarks, $\eta$, and $N_3$. 
However, its decay shows different feature depending on the scheme for
the CDM. If the CDM is $N_3$, the results shown in Fig.~2 suggest that 
the decay mode of $Z^\prime$ is mainly hadronic.  It can include 
nonhadronic ones only for the case of $\lambda~{^<_\sim}~0.33$ as
mentioned before. 
In such cases, in the $Z^\prime$ decay $\ell^+\ell^-$ + missing energy 
is also included in the final states depending on the value of $y$.
On the other hand, if one component of $\eta^0$ is the
CDM, the $Z^\prime$ always can decay into the $\eta$ pair since it
is very light. Thus, $Z^\prime$ has a substantial invisible width. 
The search of $Z^\prime$ with such features may be an important 
check of the model. 
\vspace*{10mm}

\noindent
{\Large\bf Appendix}\\
We give an example of a set of fields which cancel gauge anomalies
without affecting the discussion in the text. 
We consider to introduce additional fermions as the left-handed ones:
\begin{eqnarray}
&&2~({\bf 3}, 0, -q); \quad
3\left[({\bf 2}, +{1\over 2}, -q) + ({\bf 2}^\ast, -{1\over 2}, -q)
\right]; \quad 6\left[({\bf 1}, +1, q) + ({\bf 1}, -1, q)\right];\nonumber \\
&&5~({\bf 1}, 0, q),
\label{field}
\end{eqnarray}
where representations and charges for 
SU(2)$\times$U(1)$_Y\times$U(1)$^\prime$ are shown in parentheses.
Number of fields are also given in front of them. 
The SM gauge anomalies are canceled by taking account of these fields.
Since these fields are vector-like for the SM gauge group,
no problem is induced by them against the electroweak precision measurements. 
Although these fields are $Z_2$ odd, all of them 
can be massive through Yukawa couplings with $\phi$ or $\phi^\ast$. 
Thus, as long as their Yukawa coupling constants with $\phi$ or $\phi^\ast$ 
are simply larger than $\lambda$,
$\bar N_3$ remains as the lightest $Z_2$ odd field in the model. 
Some discrete symmetry such as $Z_2$ seems to be necessary 
to forbid the coupling between $\bar N_3$ and singlet fields shown in the
last line of (\ref{field}). However, it can be introduced without 
affecting the scenario.
Since no other seeds for the U(1)$^\prime$ breaking is 
necessary to make these additional fermions massive, 
the mass formula for $m_{Z^\prime}$ does not change and the
discussion on the relic abundance in the text is not affected.

\section*{Acknowledgement}
This work is partially supported by a Grant-in-Aid for Scientific
Research (C) from Japan Society for Promotion of Science (No.17540246).


\newpage
\begin{thebibliography}{99}
\bibitem{nmass}Super-Kamiokande Collaboration, Y.~Fukuda {\it et al.},
 Phys. Rev. Lett.{\bf 81} (1998) 1562; 
Phys. Lett. {\bf B436} (1998) 33; Phys. Lett. {\bf B433} (1998) 9;
SNO Collaboration, Q.~R.~Ahmad {\it et al.}, 
Phys. Rev. Lett. {\bf 87} (2001) 071301; Phys. Rev. Lett. 
{\bf 89} (2002) 011302; Phys. Rev. Lett. {\bf 89} (2002) 011301;
K2K Collaboration, M.~H.~Ahn {\it et al.}, 
Phys. Rev. Lett. {\bf 90} (2003) 041801; Phys. Rev. Lett. {\bf 93} 
(2004) 051801; KamLAND Collaboration, K. Eguchi {\it et al.}, 
Phys. Rev. Lett. {\bf 90} (2003) 021802 ; 
{\bf 92} (2004) 071301; T.~Araki {\it et al}, {\it ibid} {\bf 94} (2005)
081801.

\bibitem{wmap}WMAP Collaboration, D.~N.~Spergel {\it et al.},
	Astrophys. J. Suppl. {\bf 148} (2003) 175;
  SDSS Collaboration, M.~Tegmark {\it et al.}, Phys. Rev. {\bf D69}
	(2004) 103501.

\bibitem{basym}W.-M.~Yao {\it et al.}, J. phys. G {\bf 33} (2006) 1. 

\bibitem{leptg1}M.~Fukugita and T.~Yanagida, Phys. Lett. {\bf B174}
	(1986) 45.

\bibitem{leptg2}M.~Pl\"umacher, Nucl. Phys. {\bf B530} (1998) 207;
W.~Buchm\"uller and M.~Pl\"umacher, Int. J. Mod. Phys. {\bf A15} (2000)
	5047; W.~Buchm\"uller, P.~Di Bari, and M.~Pl\"umacher, Phys. Lett. 
{\bf B547} (2002) 128; Nucl. Phys. {\bf B665} (2003) 445;  
W.~Buchm\"uller, P.~Di Bari, and M.~Pl\"umacher,  Nucl. Phys. {\bf B643} 
(2002) 367; G.~F.~Giudice, A.~Notari, M.~Raidal, A.~Riotto and A.~Struma,
	Nucl. Phys. {\bf B685} (2004) 89.

\bibitem{cdm1}J.~Ellis, J.~S.~Hagelin, D.~V.~Nanopoulos, K.~A.~Olive
	and M.~Srednicki, Nucl. Phys. {\bf B238} (1984) 453;
K.~Griest, Phys. Rev. {\bf D38} (1988) 2357;
K.~Griest, M.~Kamionkowski and M.~Turner, Phys. Rev. {\bf D41} (1990) 3565;
K.~Griest and D.~Seckel, Phys. Rev. {\bf D43} (1991) 3191;
P.~Gondolo and G.~Gelmini, Nucl. Phys. {\bf B360} (1991) 145. 

\bibitem{cdm2}For recent works see for example, J
.~Ellis, K.~A.~Olive, Y.~Santoso and V.~C.~Spanos,
	Phys. Lett. {\bf B565} (2003) 176;
A.~B.~Lahanas and D.~V.~Nanopoulos, Phys. Lett. {\bf B568} (2003) 55;
H.~Baer and C.~Balazs, JCAP {\bf 0305} (2003) 006; 
U.~Chattopadhyay, A.~Corsetti and P.~Nath, Phys. Rev. {\bf D68} (2003)
	035005;
R.~Arnowitt, B.~Dutta and B.~Hu, arXiv:hep-ph/0310103;  
S.~Profumo and C.~E.~Yaguna,  arXiv:hep-ph/0407036;
E.~A.~Baltz and P.~Gondolo, arXiv:hep-ph/0407039;
G.~B\'elanger, F.~Boudjema, A.~Cottrant, A.~Pukhov and A.~Semenov, 
arXiv:hep-ph/0407218.

\bibitem{cdm2e}A.~Menon, D.~E.~Morrissey and C.~E.~M.~Wagner, Phys. Rev. {\bf
	D70} (2004) 035005;
G.~B\'elanger, F.~Boudjema, C.~Hugonie, A.~Pukhov and
	A.~Semenov, JCAP {\bf 0509} (2005) 001; 
V.~Barger, C.~Kao, P.~Langacker and H.-S.~Lee,
	Phys. Lett. {\bf B600} (2004) 104;
V.~Barger, P.~Langacker and H.-S.~Lee, Phys. Lett. {\bf B630} (2005) 85;
D.~Suematsu, Phys. Rev. {\bf D73} (2006) 035010; S.~Nakamura and 
D.~Suematsu, Phys. Rev. {\bf D75} (2007) 055004. 

\bibitem{gravitino}J.~Ellis, J.~E.~Kim and D.~V.~Nanopoulos,
	Phys. Lett. {\bf B145} (1984) 181; M.L.~Khlopov and 
A.~D.~Linde, Phys. Lett. {\bf B138} (1984) 265.

\bibitem{gravitino2}F.~Balestra, G.~Piragino, D.B.~Pontecorvo, 
M.G.~Sapozhnikov,
I.V.~Falomkin and M.Yu.~Khlopov, Sov. J. Nucl. Phys. {\bf 39} (1984) 626;
M.Yu.~Khlopov, Yu.L.~Levitan, E.V.~Sedelnikov and I.M.~Sobol,
Phys. Atom. Nucl. {\bf 57} (1994) 1393.

\bibitem{resonant}M.~Flanz, E.~A.~Pascos and U.~Sarkar, Phys. Lett. {\bf
	B345} (1995) 248; L.~Covi, E.~Roulet and F.~Vissani,
	Phys. Lett. {\bf B384} (1996) 169; A.~Pilaftsis, Phys.rev. {\bf
	D56} 1997 5431; T.~Hambye,J.~March-Russell and S.~W.~West, JHEP
	{\bf 0407} (2004) 070; A.~Pilaftsis and E.~J.~Underwood,
	Phys. Rev. {\bf D72} (2005) 113001.

\bibitem{softlept1}Y.~Grossman, T.~Kashti, Y.~Nir and E.~Roulet,
	Phys. Rev. Lett. {\bf 91} (2003) 251801;
G.~D'Ambrosio, G.~F.~Giudice and M.~Raidal, Phys. Lett. {\bf B575}
	(2003) 75; Y.~Grossman, T.~Kashti, Y.~Nir and E.~Roulet, JHEP
	{\bf 0411} (2004) 080; E.~J.~Cun, Phys. Rev. {\bf D69} (2004)
	117303; Y.~Grossman, R.~Kitano and H.~Murayama, JHEP {\bf 0506}
	(2005) 058.

\bibitem{sgravitino}T.~Asaka, K.~Hamaguchi, M.~Kawasaki and 
T.~Yanagida, Phys. Lett. {\bf B464} (1999) 12;  Phys. Rev. {\bf
	D61} (2000) 083512; H.~Murayama and T.~Yanagida, Phys. Lett. 
{\bf B322} (1994) 349; 
K.~Hamaguchi, H.~Murayama and T.~Yanagida, Phys. Rev. {\bf D65}
(2002) 043512; V.~N.~Senoguz and Q.~Shafi, Phys. Lett. {\bf B582} (2004) 6;
T.~Baba and D.~Suematsu, Phys. Rev. {\bf D71} 
(2005) 073005.

\bibitem{bdsol}S.~M.~Barr, R.~S.~Chivukula and E.~Farhi,
	Phys. Lett. {\bf B241} (1990) 387; D.~B.~Kaplan,
	Phys. Rev. Lett. {\bf 68} (1992) 741; L.~E.~Ib\'{a}\~{n}ez and
	F.~Quevedo, Phys. Lett. {\bf B283} (1992) 261; H.~Dreiner and
	G.~G.~Ross, Nucl. Phys. {\bf B410} (1993) 188; 
        R.~Kitano and I.~Low, Phys. Rev. {\bf D71} (2005) 023510;
        D.~Suematsu, J. Phys. {\bf G31} (2005) 445;
	Astropart. Phys. {\bf 24} (2006) 511; JCAP {\bf 0601} (2006) 026.

\bibitem{extsm}H.~Davoudiasl, R.~Kitano, T.~Li and H.~Murayama,
	Phys. Lett {\bf B609} (2005) 117; M.~Shaposhnikov and
	I.~Tkachev, Phys. Lett. {\bf B639} (2006) 414.

\bibitem{z2}E.~Ma, Phys Lett. {\bf B625} (2005) 76.

\bibitem{nonsusy1}L.~M.~Krauss, S.~Nasri and M.~Trodden, Phys. Rev. 
{\bf D67} (2003) 085002;
K.~Cheung and O.~Seto, Phys. Rev. {\bf D69} (2004) 113009.

\bibitem{nonsusy2}E.~Ma, Phys. Rev. {\bf D73} (2006) 077301; 
J.~Kubo, E.~Ma and D.~Suematsu, Phys. Lett. {\bf B642} (2006) 18.

\bibitem{nonsusyz}J.~Kubo and D.~Suematsu, Phys. Lett. {\bf B643} (2006) 336. 

\bibitem{nonsusyb}E.~Ma, Mod. Phys. Lett. {\bf A21} (2006) 1777;
	T.~Hambye, K.~Kannike, E.~Ma and M.~Raidal, Phys. Rev. {\bf D75}
	(2007) 095003.

\bibitem{bar}R.~Barbieri, L.~J.~Hall and V.~Rychkov,
	Phys. Rev. {\bf D74} (2006) 015007.

\bibitem{sterile}D.~Suematsu, Phys. Lett. {\bf B392} (1997) 413; 
Prog. Theor. Phys. {\bf 99} (1998) 483; Int. J. Mod. Phys. {\bf A15} 
(2000) 3967; Prog. Theor. Phys. {\bf 106} (2001) 587. 

\bibitem{radmass}A.~Zee, Phys. Lett. {\bf B93} (1980) 339;
	Phys. Lett. {\bf B161} (1985) 41; E.~Ma, Phys. Rev. Lett. {\bf
	81} (1998) 1171.

\bibitem{chooz}M.~Apollonio {\it et al.}, Phys. Lett. {\bf B466} (1999) 415.

\bibitem{cdm}E.~W.~Kolb and M.~S.~Turner, {\it The Early Universe} 
(Addison-Wesley, Redwood City, CA, 1990) 

\bibitem{wmap3}D.~N.~Spergel {\it et al.}, astro-ph/0603449.

\bibitem{leptph}F.~Abe {\it et al.}, (CDF Collaboration),
Phys. Rev. Lett. {\bf 82} (1999) 2038; T.~Affolder, (CDF Collaboration),
	Phys. Rev. Lett. {\bf 85} (2000) 2062.

\bibitem{direct}CDMS Collaboration, Phys. Rev. Lett. {\bf 96} (2006) 011302.

\bibitem{zexchange}L.~Lopez Honorez, E.~Nezri,
	J.~L.~Oliver and M.~H.~G.~Tytgat, JCAP {\bf 02} (2007) 028;
M.~Gustafsson, E.~Lundstr\"om, L.~Bergstr\"om and J.~Edsj\"o,
	Phys. Rev. Lett. {\bf 99} (2007) 041301.

\bibitem{proneu}D.~Suematsu, Prog. Theor. Phys. {\bf 96} (1996) 611.
\end{thebibliography}
\end{document}